\documentclass[a4paper]{jpconf}
\usepackage{graphicx}
\usepackage{epsf}
\usepackage[]{latexsym}
\usepackage{bm}
\newcommand{\be}{\begin{equation}}\newcommand{\ee}{\end{equation}}
\newcommand{\bea}{\begin{eqnarray}}\newcommand{\eea}{\end{eqnarray}}
\newcommand{\brr}{\begin{array}}\newcommand{\err}{\end{array}}
\newcommand{\bit}{\begin{itemize}}\newcommand{\eit}{\end{itemize}}
\newcommand{\ben}{\begin{enumerate}}\newcommand{\een}{\end{enumerate}}

\newcommand{\ba}{\begin{array}}
\newcommand{\ea}{\end{array}}

\def\lan{\langle}
\def\lf{\left}

\def\non{\nonumber}\def\ran{\rangle}
\def\rar{\rightarrow}
\def\ri{\right}
\def\ga{\gamma}
\def\de{\delta}
\def\te{\vartheta}
\def\la{\lambda}\def\si{\sigma}
\def\om{\omega}

\def\1{{_{1}}}\def\2{{_{2}}}

\newcommand{\ide}{1\hspace{-1mm}{\rm I}}

\def\noHe0{:\;\!\!\;\!\!:H_e(0):\;\!\!\;\!\!:}
\def\noHm0{:\;\!\!\;\!\!:H_\mu(0):\;\!\!\;\!\!:}

\def\vect#1{{\bm #1}}

\def\lan{\langle}
\def\lf{\left}

\def\non{\nonumber}
\def\ran{\rangle}
\def\rar{\rightarrow}
\def\ri{\right}

\def\ga{\gamma}
\def\de{\delta}
\def\te{\theta}
\def\la{\lambda}
\def\si{\sigma}
\def\om{\omega}

\def\1{{_{1}}}\def\2{{_{2}}}

\def\I{{_{\rm{I}}}}\def\II{{_{\rm{II}}}}
\def\A{{_{A}}}\def\B{{_{B}}}

\begin{document}
\title{A framework for dynamical generation of flavor mixing}

\author{M.~Blasone$^\dagger$,  P.~Jizba$^\ddagger$, G.~Lambiase$^\dagger$ and N.E.~Mavromatos$^\sharp$}

\address{$^\dagger$ INFN and  Universit\`{a} di Salerno,
Via Giovanni Paolo II, 132 -- 84084 Fisciano (SA), Italy}
\address{$^\ddagger$ FNSPE, Czech Technical
University in Prague, B\v{r}ehov\'{a} 7, 115 19 Praha 1, Czech Republic}

\address{$^\sharp$ Theoretical Particle Physics and Cosmology Group, Physics Department, King's College London, Strand, London WC2R 2LS, UK}

\begin{abstract}
We present a dynamical mechanism \`a la Nambu--Jona-Lasinio \cite{NJL} for the generation of masses and mixing for two interacting fermion fields.
The analysis is carried out in the framework introduced long ago by Umezawa et al. \cite{UTK}, in which mass generation is achieved via  inequivalent representations, and that we generalize to the case of two generations.
The method allows a clear identification of the vacuum structure for each physical phase, confirming previous results \cite{BV95} about the distinct physical nature of the vacuum for  fields with definite mass and fields with definite flavor.
Implications for the leptonic sector of the Standard Model are  briefly discussed.
\end{abstract}


\section{Introduction}
\vspace{2mm}

Particle mixing, and neutrino oscillations in particular~\cite{Bilenky:1978nj}, has undergone rapid development,
both theoretically and experimentally.
Neutrino oscillations are nowadays firmly established  by many experiments~\cite{neutrinoexp} involving solar,
atmospheric or reactor neutrinos and their basic properties
are resonably well understood\footnote{
An  exception could be the puzzling phenomenon
known as  “GSI-Oscillation-Anomaly” \cite{Kienle:2013kua,LambGSI}.}.
On the theoretical side this has, in turn, produced a large number of theoretical ideas~\cite{aba}  trying
to work out possible extensions  of the original Standard Model which does not accommodate  non-zero neutrino masses and mixings.
In spite of this, the  true origin of the mixing is still rather elusive, though it is generally believed that it is the result of physics
occurring at much higher energies than the electroweak scale.

In the context of  quantum field theory (QFT),
a complex vacuum structure has been found to arise in connection with flavor mixing \cite{BV95}.
In this approach, flavor states for mixed particles are consistently defined as eigenstates of the flavor charges:
from this, several results have been derived, including  exact oscillation formulas which exhibit corrections
with respect to the usual ones\cite{BHV98}.
However, one of the limitations of this approach resides in the fact that only free fields have been considered. This is
not a problem when discussing single-particle properties like neutrino oscillations, however, it is  inadequate if one wants to
explore salient dynamical aspects such as energy considerations related to flavor vacuum
or the dynamical generation mechanisms for such a condensate.

In this paper, we perform a first step towards this goal by addressing a more complete treatment of flavor mixing and its ensuing vacuum structure, in the context of a model with interacting fields: we consider a simple model with two
fields and dynamical symmetry breaking \`a la Nambu--Jona-Lasinio (NJL)~\cite{NJL}, which allows for the generation of (unequal) masses and of the corresponding mixing at one stroke\footnote{An early attempt in this direction can be found in Refs.\cite{BAV}. See
 also the general fomulation of the NJL model given in Ref.\cite{EriKlein}.}.
In this study, we take advantage of the reformulation of NJL mechanism given by Umezawa et al. in Ref.~\cite{UTK}, in which mass generation is achieved via  inequivalent representations, a built-in property of QFT.

  At this stage we should remark that a dynamical
generation of flavor mixing has been considered in a related but not identical context in Ref.~\cite{sarkar}. There, we have identified,
within a string-inspired framework, a microscopic mechanism for \emph{dynamical generation of  mixing} within the concept of the ``flavor vacuum''
of \cite{BV95} by identifying the latter with a ground state that was populated by space-time point-like brane defects (``D-particles'' in the brane-theory terminology~\cite{polchinski}). The Lorentz invariance breaking induced by the recoil of the defects, during their interaction with the
neutrino states in such a set up, was fully consistent with the corresponding violation of the symmetry by the flavor vacuum~\cite{BlaMagueijo}.
At an effective field theory level, the interactions of the neutrinos with the defects gave rise to effective contact four fermion interactions, which lead to dynamical formation of flavor-mixing condensates \` a la NJL model. It is the purpose of this paper to discuss a more general situation,  beyond specific models, where such a dynamical formation of flavor-vacuum condensates and the resulting mixing can be discussed based on the inequivalent representation properties of the QFT flavor vacuum of \cite{BV95}.

The present paper is organized as follows: we first review, closely following  Ref.~\cite{UTK},  the issue of inequivalent
representations in QFT and  the $V$-limit procedure, which are then used to discuss
the dynamical mass generation for the NJL model.
In Section~\ref{Sec.2} we extend the treatment to the case
of two fermion fields with an interaction term that allows
both for  unequal masses and flavor mixing generation. Results are discussed together with open issues. Section~\ref{Conclusions} is devoted to further speculations and conclusions.
%

\section{Inequivalent representations and dynamical mass generation\label{Sec.1}}
\vspace{2mm}

In this section we review the dynamical mass generation mechanism by  Nambu and Jona-Lasinio \cite{NJL}, as reformulated by Umezawa,
Takahashi and Kamefuchi~\cite{UTK}, in terms of inequivalent representations.

\subsection{Inequivalent representations in QFT \label{sec.1.1.a}}
\vspace{2mm}

It is well known that in QFT the vacuum is not a
trivial object: far from being ``empty'', it can have a rich
condensate structure with non-trivial topological properties and non-equivalent
quantum mechanical sectors (or phases). This complexity is due to the
fact that QFT possesses an infinite number of degrees of freedom and this allows
for the existence of different (unitarily inequivalent)
representations of the field algebra. So, in particular, for a given dynamics one
can have several Hilbert spaces, built on  inequivalent  vacua and
representing different phases of the system with generally very
different physical properties and distinct elementary excitations (see, e.g., Ref.~\cite{BJV,Umezawa,Miransky}). This situation is
drastically different from that of Quantum Mechanics, which deals with
systems with a finite number of degrees of freedom, and where typically only one Hilbert space
is admitted due to Stone--von Neumann's theorem~\cite{Haag,Weyl}.

In order to arrive at the concept of inequivalent representations let us consider a system of {\em Fermi} fields
enclosed in a finite-volume (volume $V$) box.
 Let  $|0\rangle$ be a fiducial reference vacuum state with the corresponding set of creation
and annihilation operators, $a_{\bf{k}}^r$ and $b_{\bf{k}}^r$, respectively. These satisfy the usual
Clifford algebra
\begin{eqnarray}
[a_{\bf{k}}^r, a_{\bf{l}}^{s\dag}]_{+} \ = \ [b_{\bf{k}}^r, b_{\bf{l}}^{s\dag}]_{+} \ = \ \delta_{\bf{k},\bf{l}}\delta_{rs}\, ,
\end{eqnarray}
with other anticommutators being zero. Here $r=1,2$ is the helicity index and
\begin{eqnarray}
{\bf k} \ = \ \frac{2\pi}{V^{1/3}}\ \!{\bf{n}}, \;\;\;\;\; n_1, n_2, n_3 \;\;\;{\mbox{integers}}\, .
\end{eqnarray}
The expansion for the field is:
\begin{eqnarray}\label{freefield}
&& \psi(x) \ = \ \frac{1}{\sqrt{V}} \sum_{{\bf k},r } \left[u_{{\bf k}}^r \ a_{{\bf k}}^r \ \!e^{  i{\bf k}\cdot{\bf x}}
 +v_{{\bf k}}^r \ b_{{\bf k}}^{r\dag} \ \!e^{  -i{\bf k}\cdot {\bf x}}\right],
\eea
where the spinor wavefunctions $u_{{\bf k}}^r$, $v_{{\bf k}}^r$ carry the time dependence through the factors $e^{- i \om_k t}$ and $e^{ i \om_k t}$, respectively, with $\om_{\bf k}=\sqrt{{\bf k}^2 + m^2}$.

Our interest lies in finding all possible unitary transformations of the vacuum state $|0\rangle$ that
satisfy simple (physically motivated) consistency  criteria. To this end we assume that the unitary transformation has the form
\begin{eqnarray}
G \ = \ e^{i F } \, ,
\end{eqnarray}
where $F$ is some self-adjoint operator which itself is some functional of
creation and annihilation operators. Invariance of
the vacuum state under translations and rotations
(vacuum is homogeneous and isotropic) implies that $G$ must satisfy the commutation relations
\begin{eqnarray}
[{\bf P}, G] \ = \ [{\bf J}, G] \ = \ 0\, ,
\label{II.4.a}
\end{eqnarray}
where ${\bf P}$ and ${\bf J}$ are total momentum and total angular momentum operators, respectively. Assuming further that the vacuum preserves
the total fermion charge we should also require that
\begin{eqnarray}
[{\cal{Q}}, G] \ = \ 0\,
\label{II.5.a},
\end{eqnarray}
with the charge ${\cal{Q}}$
\begin{eqnarray}
{\cal{Q}} \ = \ \sum_{{\bf k}, r} \left(a_{\bf{k}}^{r\dag}a_{\bf{k}}^r -  b_{\bf{k}}^{r\dag}b_{\bf{k}}^r\right) .
\end{eqnarray}
Since the vacuum states typically refer to asymptotic fields (in-fields) which have linear field equations, one can
restrict the attention to $F$'s that are only quadratic in the creation and annihilation operators that constitute the asymptotic fields.
With this the above conditions (\ref{II.4.a}) and  (\ref{II.5.a}) imply that  the unitary transformation $G$ can be
parametrized with only  two parameters $\vartheta_{{\bf k}}^{r}$ and $\varphi_{{\bf k}}^{r}$, namely (see, e.g. Ref.~\cite{BJV})
\begin{eqnarray}
G(\vartheta,\phi) \ = \ \exp\left[\sum_{{\bf k},{r}} \vartheta_{{\bf k}}^{r}\ \!\left(b_{-\bf{k}}^{r}a_{\bf{k}}^r
e^{-i\varphi_{{\bf k}}^{r}} -
a_{\bf{k}}^{r\dag} b_{-\bf{k}}^{r \dag }e^{i\varphi_{{\bf k}}^{r}} \right)   \right].
\end{eqnarray}
The invariance of $G$ under rotation ensures that $\vartheta$ and $\phi$ depend only on $k\equiv |{\bf k}|$. In addition,
it can be argued~\cite{UTK} that $\vartheta$ is independent on $r$ and $\varphi_k^r=- (-1)^r \varphi_k$. The explicit form of $G$ allows now to define new (quasi)particle
annihilation and creations operators as
\begin{eqnarray}
\alpha_{{\bf k}}^r  \ &=& \ G(\vartheta,\varphi) a_{\bf{k}}^r G^{\dag}(\vartheta,\varphi)\label{II.8.a}  \\[2mm]
&=& \ \cos\vartheta_k \ \! a_{\bf{k}}^r + e^{i\varphi_k^{r}} \sin \vartheta_k
\ \! b_{-\bf{k}}^{r \dag }\, , \label{II.8.abc}  \\[3mm]
\beta_{{\bf k}}^r  \ &=& \ G(\vartheta,\varphi)b_{\bf{k}}^r G^{\dag}(\vartheta,\varphi) \label{II.8.ac}\\[2mm]
&=& \ \cos \vartheta_k \ \! b_{\bf{k}}^r -
e^{i\varphi_k^{r}} \sin \vartheta_k  \ \! a_{-\bf{k}}^{r\dag}\, .\label{II.8.aa}
\end{eqnarray}
The corresponding inverse transformation can be easily deduced:
\begin{eqnarray}
a_{\bf{k}}^r \ &=& \  \ G^{\dag}(\vartheta,\varphi) \alpha_{\bf{k}}^r G(\vartheta,\varphi)
\ = \ \cos\vartheta_k \ \! \alpha_{\bf{k}}^r - e^{i\varphi_k^{r}} \sin \vartheta_k
\ \! \beta_{-\bf{k}}^{r \dag }\, ,  \nonumber \\[2mm]
b_{\bf{k}}^r \ &=& \  \ G^{\dag}(\vartheta,\varphi) \beta_{\bf{k}}^r G(\vartheta,\varphi)
\ = \ \cos \vartheta_k \ \! \beta_{\bf{k}}^r +
e^{i\varphi_k^{r}} \sin \vartheta_k  \ \! \alpha_{-\bf{k}}^{r\dag}\, . \label{II.8.ab}
\end{eqnarray}
The transformations (\ref{II.8.a})-(\ref{II.8.ab}) preserve the commutation relations, and represent thus a
Bogoliubov transformation in the usual sense~\cite{BJV}. In this respect the labels $\{\vartheta_k,
\varphi_k^{r}\}$ yield the most general  parametrization for the Bogoliubov transformation of creation
and annihilation operators. A physical picture behind the Bogoliubov transformation (\ref{II.8.aa})
is that $a_{\bf{k}}^{r\dagger}$ (and $b_{\bf{k}}^{r\dagger}$) create above the vacuum state $|0\rangle$ particle (anti-particle)
quanta with momentum ${\bf{k}}$ and helicity $r$, whereas the quasiparticles (and anti-quasiparticles) created
by $\alpha_{{\bf k}}^{r\dagger}$ (and $\beta_{{\bf k}}^{r\dagger}$) are the elementary excitations above the vacuum state
\begin{eqnarray}
|0(\vartheta,\varphi)\rangle \ = \ G(\vartheta,\varphi) |0\rangle  \ = \ \prod_{{\bf k},{r}}\left(\cos\vartheta_k \ - \ e^{i\varphi_k^{r}}
\sin \vartheta_k \ \! a_{\bf{k}}^{r\dagger} b_{-\bf{k}}^{r\dagger}   \right)\!|0\rangle\, ,
\label{II.10.aa}
\end{eqnarray}
which is annihilated both by $\alpha_{{\bf k}}^{r}$ and $\beta_{{\bf k}}^{r}$.

In the finite volume limit all vacuum states $|0(\vartheta,\varphi)\rangle$ are   equivalent
(i.e., they describe the same unique physical ground-state).
In the infinite-volume limit the situation is drastically different.  This can be seen by noticing that from (\ref{II.10.aa}) we have (for $V\rar \infty$):
\begin{eqnarray}
\langle 0| 0(\vartheta,\varphi)\rangle \ = \ \exp\left[\sum_{{\bf k}, r} \log(\sin \vartheta_k) \right] \ = \
\exp\left[\frac{V}{(2\pi)^3} \int d^3 {\bf k} \log(\sin^2 \vartheta_k) \right]
\
\rightarrow \  0\, .
\end{eqnarray}
Since as the fiducial vacuum one can chose any of the infinitely many vacuum states $| 0(\vartheta,\varphi)\rangle$, the previous result
implies that in the infinite-volume limit all the vacua with different $\vartheta$'s and $\varphi$'s are orthogonal, i.e.,
\begin{eqnarray}
\langle  0(\vartheta,\varphi)| 0(\vartheta',\varphi') \rangle \  \rightarrow \  0\, , \;\;\;\;\;\; \vartheta',\varphi' \neq\ \vartheta,
\varphi\, .
\end{eqnarray}
%
%
 The representation (\ref{II.8.a}) (or (\ref{II.8.ac})) of the Bogoliubov transformation
loses its meaning for an (infinite)
QFT system, in as much as the operator $\exp(iF)$ occurring in it has no domain on the
representation space involved~\cite{Umezawa}. This fact, however, has no direct bearing on QFT
which uses the well-defined form (\ref{II.8.abc}) (or (\ref{II.8.aa})). It just states that the operator-algebra
representations which are used in QFT yield unitarily
inequivalent Fock-space representations. In other words, the vacuum states $|0(\vartheta',\varphi') \rangle$ and
$|0(\vartheta,\varphi) \rangle$ for different $\vartheta$'s and $\varphi$'s do not belong to the same Hilbert space.

Broken symmetry is a typical framework in which the above inequivalent-representations picture is of a
particular importance.  This is because the different vacuum states describing the broken symmetry phases
cannot be connected by unitary
representatives of the symmetry group in question~\cite{Emch}. What happens there is that each vacuum state
induces a truly different representation of the operator algebra in
each broken phase~\cite{BJV,Umezawa}.
Such a multiple vacuum structure was used in Ref.~\cite{UTK} to reformulate mass generation via dynamical breakdown of symmetry  in the Nambu--Jona-Lasinio model. We will review this  in Section~\ref{sec.1.3.a} and further extend in Section~\ref{Sec.2} to the case of two generations in order to accomodate for flavor mixing.

%

\subsection{$V$-limit procedure\label{sec.1.2.a}}
\vspace{2mm}

To proceed, we review now the so-called $V$-limit procedure introduced by Umezawa {\em et al.} in Ref.~\cite{UTK}.
Let us consider
matrix elements of QFT operators, say $Q$, between states $|\Phi_i(\vartheta,\varphi)\rangle$, generated from the vacuum state $|0(\vartheta,\varphi)\rangle$
by a suitable action of creation and annihilation operators. The index ``$i$" is a multi-index distinguishing various states, and the two real parameters $\vartheta$ and $\varphi$ label the different (unitarily inequivalent) vacuum states.
In particular the $V$-limit of $Q$ with respect to a representation characterized by the parameters $\{\vartheta,\varphi\}$
is defined as
\begin{eqnarray}
&&\langle \Phi_i(\vartheta,\varphi)| \mbox{$V$-lim}[Q]|\Phi_j(\vartheta,\varphi)\rangle \equiv \ \lim_{V\rightarrow \infty}\langle \Phi_i(\vartheta,\varphi)| Q|\Phi_j(\vartheta,\varphi)\rangle\, ,
\label{Eq.5a}
\end{eqnarray}
for all $i$ and $j$. It should be noted that the $V$-limit is not the same as the week limit because the basis of the representation
in which the limit
is carried out may depend (and as a rule it does) on the volume $V$. The matrix element on the right-hand side of (\ref{Eq.5a}) is operationally calculated
by phrasing the full (Heisenberg-picture) fields $\psi$ present in $Q$ in terms of free fields $\psi_{\rm in}$ enclosed in a finite-volume (volume $V$) box. The mapping between $\psi$ and $\psi_{\rm in}$  is known as the Yang--Feldman equation or also Haag's map~\cite{BJV,Yang-Feldman,Bog,BJ02AP}. Formally it can be written in the form~\cite{BJV,Bog}; $\psi(x) = S^\dag T(S \psi_{\rm in}(x))$, where $S$ and $T$ are the $S$-matrix and time-ordering symbol, respectively.

The free field in the representation $\{\vartheta,\varphi \}$ can be obtained from the  free field expansion (\ref{freefield}) via the Bogoliubov transformation (\ref{II.8.a})-(\ref{II.8.aa}). As a result one has, for the {\em same} field operator:
\begin{eqnarray}
&&
\psi (x) \ = \ \frac{1}{\sqrt{V}} \sum_{{\bf k},r } \left[u_{{\bf k}}^r(\vartheta,\phi) \alpha_{{\bf k}}^r \ \!e^{  i{\bf k}\cdot{\bf x}}
 +v_{{\bf k}}^r(\vartheta,\phi) \beta_{{\bf k}}^{r\dag} \ \!e^{  -i{\bf k}\cdot {\bf x}}\right],
\end{eqnarray}
with
\begin{eqnarray}
\alpha_{{\bf k}}^r\, |0(\vartheta,\varphi)\rangle \ = \ \beta_{{\bf k}}^{r}\,  |0(\vartheta,\varphi)\rangle \ = \ 0\, .
\end{eqnarray}
The Dirac spinors $u_{{\bf k}}^r(\vartheta,\phi)$ and $v_{{\bf k}}^r(\vartheta,\phi)$ are related with
the fiducial representation spinors via the relation
\begin{eqnarray}
&&u_{{\bf k}}^r(\vartheta,\phi) \ = \ u_{{\bf k}}^r \ \!  \cos \vartheta_k \ + \ v_{-{\bf k}}^r \ \! e^{-i\varphi_k^{r}} \sin \vartheta_k\, ,\nonumber \\[2mm]
&&v_{{\bf k}}^r(\vartheta,\phi) \ = \ v_{{\bf k}}^r \ \! \cos \vartheta_k \ - \ u_{-{\bf k}}^r \ \! e^{i\varphi_k^{r}}\sin \vartheta_k\, .
\end{eqnarray}

By employing the operatorial Wick theorem~\cite{Bogoliubov}, it is a simple exercise to show that for
{\em free} fields   in the $\{\vartheta,\varphi \}$ representation we have
\begin{eqnarray}
&&{\mbox{$V$-lim}}\left[\int d^3{\bm{x}}\ \! \bar{\psi}_{\alpha}(x)\psi_{\beta}(x)\right] \ = \
\int d^3{\bm{x}}\ \!
:\bar{\psi}_{\alpha}(x)\psi_{\beta}(x): \ + \ \int d^3{\bm{x}}\ \! iS^{-}_{\alpha\beta}(\vartheta,\varphi)\, ,\nonumber \\[3mm]
&&{\mbox{$V$-lim}}\left[\int d^3{\bm{x}}\ \! \bar{\psi}_{\alpha}(x)\psi_{\beta}(x) \bar{\psi}_{\gamma}(x)\psi_{\delta}(x)  \right] \, =\,\nonumber \\[2mm]
&&\mbox{\hspace{30mm}}= \ iS^{-}_{\alpha\beta}(\vartheta,\varphi)\int d^3{\bm{x}}\ \!
:\bar{\psi}_{\gamma}(x)\psi_{\delta}(x): \ + \ iS^{+}_{\gamma\delta}(\vartheta,\varphi)\int d^3{\bm{x}}\ \!
:\bar{\psi}_{\alpha}(x)\psi_{\beta}(x): \nonumber \\[2mm]
&&\mbox{\hspace{30mm}}+ \ \ iS^{-}_{\alpha\delta}(\vartheta,\varphi)\int d^3{\bm{x}}\ \!
:\bar{\psi}_{\gamma}(x)\psi_{\beta}(x): \ + \ iS^{+}_{\gamma\beta}(\vartheta,\varphi)\int d^3{\bm{x}}\ \!
:\bar{\psi}_{\alpha}(x)\psi_{\delta}(x): \nonumber \\[2mm]
&&\mbox{\hspace{30mm}}+ \int d^3{\bm{x}}\ \! \sum_{\rm{contractions}} S^{+}(\vartheta,\varphi)S^{+}(\vartheta,\varphi)\, .
\label{Eq.9b}
\end{eqnarray}
where we have introduced the  two-point Wightman functions evaluated with respect to the $|0(\vartheta,\varphi)\rangle $ vacuum:
\begin{eqnarray}
&&iS^{+}_{\alpha\beta}(\vartheta,\varphi)\ = \
\langle 0(\vartheta,\varphi)| \bar{\psi}_{\alpha}(x)\psi_{\beta}(x)|0(\vartheta,\varphi)\rangle ,
\\ [2mm]
&& i{S}^{-}_{\beta\alpha} (\vartheta,\varphi)\ = \  \langle 0(\vartheta,\varphi)|\psi_{\beta}(x)\bar{\psi}_{\alpha}(x) |0(\vartheta,\varphi)\rangle\, .
\end{eqnarray}

Note that, due to translational invariance of the vacuum state $|0(\vartheta,\varphi)\rangle$, the two-point Wightman function is
$x$-independent. This implies that the last terms in  both equations in (\ref{Eq.9b}) are the c-numbers
proportional to the volume $V$.

It is useful to consider the explicit form of the following quantities (see Ref.\cite{UTK}):
\begin{eqnarray} \label{CpCs}
C_p&\equiv & i \lim_{V\rightarrow \infty}\,
\langle 0(\vartheta,\varphi)| \bar{\psi}(x) \ga_5 \psi(x)|0(\vartheta,\varphi)\rangle \
=\ \frac{2}{(2 \pi)^3} \int d^3{\bf k}\, \sin 2\vartheta_k \ \sin \varphi_k
\\ \non
C_s&\equiv & \lim_{V\rightarrow \infty}\,
\langle 0(\vartheta,\varphi)| \bar{\psi}(x)  \psi(x)|0(\vartheta,\varphi)\rangle \
=\ - \frac{2}{(2 \pi)^3} \int d^3{\bf k}\, \lf[ \frac{m}{\om_k} \cos 2\vartheta_k \,
 - \frac{k}{\om_k} \sin 2\vartheta_k \cos \varphi_k \ri].
\eea

\subsection{Gap equation \label{sec.1.3.a}}
\vspace{2mm}

Following Ref.\cite{UTK}, we now apply the above developed concepts to the study of mass generation in the NJL model. This is described by
the following Hamiltonian\footnote{Here, as in Ref.\cite{UTK}, we consider the general case where $m\neq 0$. Results for $m=0$ are then obtained as a special case.}
\begin{eqnarray}
&&{H} \ = \ {H}_0 + {H}_{\rm{int}}\, , \\[2mm]
&&{H}_0 \ = \ \int d^3{\bf{x}} \ \! \bar{{{\psi}}}  \left( -i\vect{\gamma}\cdot\!\vect{\nabla}\ + \  m\right) {{\psi}}\, ,
  \\[1mm]
&&{H}_{\rm{int}} \ = \ \lambda \int d^3 {\bf{x}} \ \! \left[\left(\bar{{{\psi}}} {{\psi}}\right)^2  -
\left(\bar{{{\psi}}} \gamma^5{{\psi}}\right)^2\right]\, .
\label{Eq.3.1}
\end{eqnarray}
Considering the lowest order in the Yang--Feldman expansion, the $V$-limit of $H$ gives \cite{UTK}:
\begin{eqnarray}
 V\mbox{-lim}\left[{H}\right] \ &=& \  \bar{ H}_0
 \,  + \, \mbox{c-number}\, ,
\label{Eq.3.2}
\end{eqnarray}
with
\begin{eqnarray}
&& \bar{H}_0\, =\, H_0 + \delta H_0\, , \qquad
\delta H_0 \  =\  \int d^3{\bf{x}} \ \lf\{ f \, \bar{\psi} \psi  \ + \ i  g \, \bar{\psi} \ga_5 \psi \ri\}.
\end{eqnarray}
and
\be \label{fgC}
f \, = \, \la\, C_s \, ,\qquad  g \, = \, \la\, C_p\, .
\ee

So far we have not specified the vacuum state $|0(\vartheta,\varphi)\rangle$ of interest. In fact, most of the vacuum
states and ensuing representations are not physically acceptable. The physically admissible representations are only
those which
satisfy appropriate renormalization condition, namely that  the $V$-limit of the full Hamiltonian $H$
should describe the quasiparticle (i.e., diagonal) Hamiltonian with the correct relativistic dispersion condition. Thus one requires that \cite{UTK}:
\begin{eqnarray} \label{H0diag}
\bar{H}_0  \ = \ \sum_{r} E_{k}\left(\alpha_{{\bf k}}^{r\dag}\alpha_{{\bf k}}^r +
\beta_{{\bf k}}^{r\dag}\beta_{{\bf k}}^r \right) \ + \ W_{0}\, ,
\end{eqnarray}
with the free particle dispersion relation $E_k = \sqrt{k^2 + M^2}$. The mass $M$ corresponds to the mass of elementary excitations
(or quasiparticles) over the physical vacuum. The vacuum energy (or condensate density $W_0$) is fixed by setting the values of  parameters $\vartheta$ and $\varphi$ of the physical  representation into the expression \cite{UTK}:
\bea \non
W_0&=& \sum_r \int d^3{\bf{k}} \ \lf[ 2 \om_k \sin^2\vartheta_k - (-1)^r g \sin 2 \vartheta_k \sin\varphi_k^r \ri.
\\ \label{W0}
&&\lf. \hspace{1cm}
- \frac{f}{\om_k}\lf(m \cos 2\vartheta_k - k \sin 2\vartheta_k \cos\varphi_k^r\ri) - \om_k\ri]
\, = \, -2 \int d^3{\bf{k}} \, E_k \,.
\eea

After tedious but straightforward calculations  one finds that the condition (\ref{H0diag}) is satisfied\footnote{The condition $E_k>0$ is also enforced.} when the
following conditions hold~\cite{UTK}
\begin{eqnarray}
\cos(2\vartheta_k) & = &
\frac{1}{E_{k}}\left[\frac{m}{\omega_{k}} f(\vartheta, \varphi) +
\omega_{k}\right], \label{MGE1a} \\[2mm]
\sin(\varphi_{k}^r) & = & g(\vartheta, \varphi)(-1)^r  \ \!\left[g^2(\vartheta, \varphi) +
\frac{{k}^2}{\omega_{k}^2} f^2(\vartheta, \varphi)\right]^{-\frac{1}{2}}, \label{MGE1b}  \\[2mm]
M^2(\vartheta, \varphi)  & = & (m \ + \ f(\vartheta, \varphi))^2 + g^2(\vartheta, \varphi)
\ = \  (m \ + \ \la C_s)^2 + \la^2 C_p^{\,2}\, ,
\label{MGE1}
\end{eqnarray}

Since $f$ and $g$ depend on the parameters $\{\vartheta, \varphi\}$ via Eqs.(\ref{fgC}) and (\ref{CpCs}), the above solutions give rise to  two non-linear equations
\begin{eqnarray}
f \ = \ f(\vartheta(f,g), \varphi(f,g))\;\;\;\; \mbox{and} \;\;\;\; g \ = \ g(\vartheta(f,g), \varphi(f,g))\, ,
\label{eq-39}
\end{eqnarray}
which can be recasted as
\bea
C_p \lf(1 + \frac{2\la}{(2\pi)^3}\int \frac{d^3{\bf k}}{E_k}\ri) &=&0\,,
\\
C_s \lf(1 + \frac{2\la}{(2\pi)^3}\int \frac{d^3{\bf k}}{E_k}\ri) &=& - \frac{2 m}{(2\pi)^3}\int \frac{d^3{\bf k}}{E_k}\,.
\eea
These equations determine the mass M.

In Ref.~\cite{UTK} two possibilities are discussed
\bea \label{gapeqA}
C_p = 0, & & M = m \,-\, \frac{2\la }{(2\pi)^3}\, M \int \frac{d^3{\bf k}}{E_k},
\\ \label{gapeqB}
m = 0, & & 1 + \frac{2\la}{(2\pi)^3}\int \frac{d^3{\bf k}}{E_k} \, =\,0.
\eea
The second case, Eq.~(\ref{gapeqB}), is only allowed for $\la <0 $. Eq.~(\ref{gapeqA}) for $m=0$ and $M\neq 0$ is a special case of Eq.~(\ref{gapeqB}).

For $m\neq 0$, Eq.~(\ref{gapeqA}) gives perturbative corrections to the mass:
\bea
M &=& m \, - \, \frac{2\la }{(2\pi)^3}\, m \int \frac{d^3{\bf k}}{\om_k} \, + \, \cdots
\eea

On the other hand, the solution Eq.~(\ref{gapeqB}) has a non-perturbative character and expresses the dynamical breakdown of (chiral) symmetry.

%

\section{Two-flavor mixing\label{Sec.2}}
\vspace{2mm}

We now consider the dynamical symmetry breaking for the case of two fermion fields, for which in general a non-diagonal mass matrix will be obtained, thus generating flavor mixing in addition to nonzero masses. Here we mainly intend to present some qualitative aspects of this extension, relegating a complete discussion to a future paper. The notation is over-simplified:  we  omit spacetime dependence and drop momentum and helicity indices.
Let us consider a fermion field
doublet ${\vect{\psi}}$ whose Hamiltonian density is given as
\begin{eqnarray}\label{generalH}
&&{\cal H} \ = \ {\cal H}_0 + {\cal H}_{\rm{int}}\, ,\\[2mm]
&&{\cal H}_0 \ = \  {\bar{\vect{\psi}}} \left(
-i\vect{\gamma}\cdot\!\vect{\nabla}\ + \ \bm{M}_0\right)
 {\vect{\psi}}\, ,
\label{Eq.1}
\end{eqnarray}
with $\vect{\gamma}$ being a shorthand for $\ide\otimes\vect{\gamma}$ and
\begin{eqnarray}
{\vect{\psi}} \ = \ \left(
                       \begin{array}{c}
                         \psi_{_{\rm{I}}} \\
                         \psi_{_{\rm{II}}} \\
                       \end{array}
                     \right) \;\;\;\;\; \mbox{and} \;\;\;\;\; \bm{M}_0 \ =
                     \ \left( \begin{array}{cc}m_{_{\rm{I}}} & {0}\\{0}
& m_{_{\rm{II}}}
                     \\ \end{array} \right)\otimes \ide\, ,
\label{Eq.2}
\end{eqnarray}
where $\ide$ is the $2\times 2$ identity matrix.
The interaction Hamiltonian ${H}_{\rm{int}}$ can be assumed in the generic form
\begin{eqnarray}\label{Hintfull}
{\cal H}_{\rm{int}} \ = \
\left(\bar{{\vect{\psi}}}\,\Gamma \,
{\vect{\psi}}\right)\left(\bar{{\vect{\psi}}}\, \Gamma' \, {\vect{\psi}}\right),
\end{eqnarray}
where $\Gamma$ and $\Gamma'$ are some doublet spinor matrices.
For simplicity we will, in the following, consider only the scalar counterterms, i.e.,
we will put $g_\I = g_\II =0$. Recalling Eq.(\ref{MGE1b}), this in turn implies that
$\varphi_\I=\varphi_\II=0$ in the Bogoliubov transformations for fields
$\psi_\I$ and $\psi_\II$. This assumption simplifies considerably the
following treatment, without altering the results of our
analysis. The case including also pseudoscalar potential and related
counterterms will be discussed elsewhere.

In this case the $V$-limit renormalization term $\delta \mathcal{H}_0$ has the generic structure
%
%
\begin{eqnarray}
\delta \mathcal{H}_0  &=&  \delta \mathcal{H}_0^\I \, +\, \delta
\mathcal{H}_0^\II\, +\,
\delta \mathcal{H}_{\rm{mix}}\\[2mm]
&=& f_\I \,\bar{\psi}_\I \psi_\I
\ + \ f_\II\, \bar{\psi}_\II \psi_\II \ + \
 h  \, \left(\bar{\psi}_\I \psi_\II \ + \
{\bar \psi}_\II \psi_\I \right)\, .
\end{eqnarray}
%

Instead of the Bogoliubov transformation (\ref{II.8.ab}),
inequivalent representations are now defined through
a $4\times 4$ canonical transformation, which can be conveniently
parameterized as:
\bea
  \left( \begin{tabular}{c} $\alpha_A$ \\ $\beta_A^\dagger$
\\$\alpha_{B}$ \\ $\beta_{B}^\dagger$ \end{tabular} \right)
= \left(\begin{array}{cccc}
c_\theta\, \rho_{\A\I}& c_\theta \,\lambda_{\A\I} &
s_\theta \,\rho_{\A\II}  &
s_\theta \,\lambda_{\A\II}
\\  - c_\theta \,\lambda_{\A\I} &c_\theta\, \rho_{\A\I} & - s_\theta
\,\lambda_{\A\II} & s_\theta \,\rho_{\A\II}
\\  - s_\theta \,\rho_{\B \I} &-  s_\theta \,\lambda_{\B \I}& c_\theta
\,\rho_{\B \II}
&  c_\theta \,\lambda_{\B \II} \\   s_\theta \,\lambda_{\B \I} &-
s_\theta\,
\rho_{\B \I} & - c_\theta\, \lambda_{\B \II}& c_\theta\, \rho_{\B \II}
\end{array}\right)
  \left( \begin{tabular}{c} $a_\I$ \\ $b_\I^\dagger$ \\  $a_\II$
\\$b_\II^\dagger$ \end{tabular} \right).
  \label{4x4Bog}
  \eea
where $c_\theta\equiv \cos\theta$, $s_\theta\equiv \sin\theta$ and
\begin{eqnarray}\label{rholambda}
  \rho_{a b} \ \equiv \ \cos\frac{\chi_a - \chi_b}{2}, \quad
\lambda_{a b} \ \equiv \  \sin\frac{\chi_a -
\chi_b}{2} \, ,
\quad
\chi_a \ \equiv \  \cot^{-1}\lf[\frac{k}{m_a}\ri], \qquad a,b={\rm I},{\rm II},A,B\, .
\end{eqnarray}
The transformation~(\ref{4x4Bog}) contains thus three parameters
$(\theta, m_A, m_B)$ to be fixed in
terms of the quantities ($f_\I,f_\II,h$) in order to diagonalize the
Hamiltonian.

Let us start by considering the case in which no mixing arises after the
$V$-limit. In such a situation, the Hamiltonian reduces into the sum
of two singlet-field Hamiltonians, each being the same as the one studied
in the previous section, i.e.
\be
  {\bar {\cal H}}_0 \ = \ \sum_{i={\rm I},{\rm II}}\lf({\cal H}_0^i  \ + \  \delta \mathcal{H}_0^i  \ri).
  \label{48a}
\ee
The Bogoliubov matrix $(\ref{4x4Bog})$ that describes this situation must be block diagonal, i.e.
\bea
  \left( \begin{tabular}{c} $\alpha_A$ \\ $\beta_A^\dagger$
\\
$\alpha_{B}$ \\ $\beta_{B}^\dagger$ \end{tabular} \right)
\ = \
  \left(\begin{array}{cccc}   \rho_{\A\I}&  \lambda_{\A\I} &
0 &0
\\- \lambda_{\A\I} &  \rho_{\A\I} &  0 & 0
\\ 0 &0& \rho_{\B \II}
&  \lambda_{\B \II} \\ 0 &0 &   -\lambda_{\B \II}& \rho_{\B \II}
\end{array}\right)
  \left( \begin{tabular}{c} $a_\I$ \\ $b_\I^\dagger$ \\ $a_\II$
\\$b_\II^\dagger$ \end{tabular} \right),
  \eea
and the diagonalization condition reads (cf Eq.(\ref{MGE1})):
\bea
  m_A \ = \ m_\I+f_\I\,, \quad m_B \ = \  m_\II+f_\II\, . \label{substUTK}
\eea
This is the same as the condition derived in Section~\ref{sec.1.3.a}, if
we make the identification
\be
\vartheta_i = \frac{1}{2}\lf(\cot^{-1}\lf[\frac{k }{m_a}\ri] -
\cot^{-1}\lf[\frac{k }{m_i}\ri]\ri), \qquad  (a,i)=(A,{\rm I}),(B,{\rm II})\, .
\ee
The resulting  Hamiltonian (\ref{48a}) is now expressed in terms of the $A,B$ modes.

\medskip

Let us now come back to the full Hamiltonian (\ref{generalH}). After the $V$-limit, in general  we obtain
an Hamiltonian density of the form:
  \be\label{HfullV}
  {\bar {\cal H}}_0 \ = \ \sum_{i={\rm I},{\rm II}}\lf({\cal H}_0^i \ + \ \delta \mathcal{H}_0^i \ \ri)
\ + \ \de {\cal H}_{\rm{mix}}\, .
  \ee

In order to select among the inequivalent representations, we have to impose an appropriate renormalization condition on the
form of the Hamiltonian (\ref{HfullV}). With respect to the simple case described in
Section~\ref{sec.1.3.a}, where only one field
was present, we have now two distinct possibilities:

\begin{itemize}
\item
One possibility is to impose the condition that the Hamiltonian
(\ref{HfullV}) becomes fully diagonal in two fermion
fields, $\psi_1$ and $\psi_2$, with masses $m_1$ and $m_2$:
\be\label{H12}
{\bar {\cal H}}_0  \ = \ \sum_{j=1,2}  \bar{\psi}_j  \left(
-i\vect{\gamma}\cdot\!\vect{\nabla}\ + \  m_j\right)  {{\psi_j}}\, .
\ee

The condition for the complete diagonalization of (\ref{HfullV}) is
found to be:
\begin{eqnarray}
\theta &\rightarrow& \bar{\theta} \ \equiv \
\frac{1}{2}\, \tan^{-1}\left[\frac{2h}{m_{\mu}-m_{e}}\right], \label{fullmixdiag-a} \\[2mm]
m_\A & \rightarrow & m_1 \ \equiv \
\frac{1}{2}\left(m_{e}+m_{\mu}-\sqrt{(m_{\mu}-m_{e})^2+4h^2}\right),
\label{fullmixdiag-b}\\[2mm] \label{fullmixdiag}
m_\B & \rightarrow & m_2 \ \equiv \
\frac{1}{2}\left(m_{e}+m_{\mu}+\sqrt{(m_{\mu}-m_{e})^2+4h^2}\right).
\end{eqnarray}
where we introduced the notation $m_{e} = m_\I + f_\I$, $m_{\mu} = m_\II + f_\II$.
In passing we might observe that (\ref{fullmixdiag-a})-(\ref{fullmixdiag}) imply the useful mass relations
\begin{eqnarray}
m_e \ &=& \ m_1 \cos^2\bar{\theta} \ + \ m_2 \sin^2\bar{\theta}\nonumber \\[2mm]
m_\mu \ &=& \ m_2 \cos^2\bar{\theta} \ + \ m_1 \sin^2\bar{\theta}\, .\label{mass-rel}
\end{eqnarray}
In the following we will denote the vacuum state associated with such a representation as
\begin{eqnarray}
|0(\bar{\theta}, m_1, m_2) \rangle \ \equiv \ |0\ran_{1,2}\, ,
\end{eqnarray}
since it is simply the
tensor product states of the vacua for the free fields $\psi_1$ and
$\psi_2$. In addition, the vacuum expectation value ${}_{1,2}\lan 0 |\cdots| 0 \ran_{1,2}$
will be denoted for short as $\langle \cdots \rangle_{1,2}$. With
this the vacuum expectation value of the Hamiltonian in this representation
can be shown to have the form:
\be\label{vevHprime}
  \lan {\bar H}_0  \ran_{1,2} \ = \ - 2 \int d^3{\bf{k}} \
\left(\sqrt{k^2\ + \ m_1^2} \ + \ \sqrt{k^2 \ + \ m_2^2}\right) .
\ee

\item
Another possible representation is obtained by a partial diagonalization
of (\ref{HfullV}), leaving untouched $ \de {\cal H}_{\rm{mix}}$. This will lead to the Hamiltonian density
  \be\label{Hem}
{\bar \mathcal{H}}_0  \ = \ \sum_{\si=e,\mu}    \bar{\psi}_\si \left(
-i\vect{\gamma}\cdot\!\vect{\nabla}\ + \  m_\si\right) {{\psi_\si}}
\,+\,
  h\,  ({\bar \psi}_e \psi_\mu \ + \ {\bar \psi}_\mu \psi_{e})\, .
\ee
Such a representation is obtained by setting
  \bea
\te &\rar& 0\, ,\\
m_A &\rar & m_e \ \equiv \ m_\I+f_\I\, ,
\\[2mm]
m_B &\rar & m_\mu \ \equiv \ m_\II+f_\II\,.
  \eea
The vacuum in this representation is denoted as
\begin{eqnarray}
|0(\theta=0, m_e, m_\mu )\rangle \ \equiv \ |0\rangle_{e\mu}
\, ,
\end{eqnarray}
and will be called the flavor vacuum. An important point to be noticed is that the mixing term in
Eq.(\ref{Hem}) is form-invariant under the transformation
(\ref{4x4Bog}), provided $\te=0$.
\vspace{2mm}

Denoting $ {}_{e,\mu}\lan 0 |\cdots | 0 \ran_{e,\mu}$ as $\langle \cdots \rangle_{e,\mu}$, the vacuum expectation value of the Hamiltonian in this representation
is given as:
\be\label{vevHprimeb}
  \lan {\bar H}_0 \ran_{e,\mu}  \ = \ - 2 \int d^3{\bf{k}} \
\left(\sqrt{k^2\ + \ m_e^2} \ + \ \sqrt{k^2 \ + \ m_\mu^2}\right) ,
  \ee
since $\lan \de {\cal H}_{\rm mix}  \ran_{e,\mu}  \ = \ 0$.
\end{itemize}

In passing it should be stressed that the expressions for vacuum energies (\ref{vevHprime}) and (\ref{vevHprimeb})
should have some ultraviolet momentum cutoff in order to ensure meaningful stable vacua.

\subsection{Physical motivations for the choice of the representation}
\vspace{2mm}

The choice between the representations $|0\ran_{1,2}$ and $| 0 \ran_{e,\mu}$ has to be motivated on physical grounds.
In this sense, the requirement that the Hamiltonian is only partially diagonalized, cf. Eq.(\ref{Hem}), which corresponds to the representation built on the flavor vacuum  $|0 \ran_{e,\mu}$, seems to be the one which better fits the situation present in the Standard Model, where the flavor fields describe the physical particles, and do not have in general a diagonal mass matrix\footnote{The fact that charged leptons, for example, are not mixed, while neutrinos are, is just a matter of convention, since the generation of masses via the Higgs mechanism produces non-diagonal mass matrices for all fermions.}.
In this representation, mixing can be seen as the effect of an external field, as discussed in Refs.\cite{Blasone:2010zn}.

The difference between the two above representations can be also seen via the respective gap equations. In particular, in the representation $|0\ran_{1,2}$ the gap equation will be formally written as
a set of $3$ non-linear equations for $f_\I$, $f_\II$ and $h$ in the form
\begin{eqnarray}
&&f_\I \ = \ f_\I(\bar{\theta},m_1,m_2) \ = \
f_\I(\bar{\theta}(f_\I,f_\II,h), m_1(f_\I,f_\II,h), m_2(f_\I,f_\II,h))\, , \nonumber \\[2mm]
&&f_\II \ = \ f_\II(\bar{\theta},m_1,m_2) \ = \  f_\II(\bar{\theta}(f_\I,f_\II,h), m_1(f_\I,f_\II,h), m_2(f_\I,f_\II,h))\, ,\nonumber \\[2mm]   &&h \ = \ h(\bar{\theta},m_1,m_2) \ = \ h(\bar{\theta}(f_\I,f_\II,h), m_1(f_\I,f_\II,h), m_2(f_\I,f_\II,h))\, .
\end{eqnarray}
Here the explicit forms of $f_\I$, $f_\II$ and $h$ are determined in terms of the vacuum expectation values  $\langle \bar{\psi}_\I \psi_\I \rangle_{1,2}$, $\langle \bar{\psi}_\I \psi_\II \rangle_{1,2}$ and $\langle \bar{\psi}_\II \psi_\II \rangle_{1,2}$. This is a direct generalization of Eq.(\ref{eq-39}).

On the other hand, in the representation $|0 \ran_{e,\mu}$, the relevant non-linear equations to be solved are
\begin{eqnarray}
\tilde{f} \ = \ \tilde{f}(m_e, m_{\mu}) \ = \ \tilde{f}(m_e(\tilde{f},\tilde{h}), m_{\mu}(\tilde{f},\tilde{h}))\, , \nonumber \\[2mm]
\tilde{h} \ = \ \tilde{h}(m_e, m_{\mu}) \ = \ \tilde{h}(m_e(\tilde{f},\tilde{h}), m_{\mu}(\tilde{f},\tilde{h})) \, .
\end{eqnarray}
Again, the explicit forms of $\tilde{f}$ and $\tilde{h}$ are determined from the expectation values
$\langle \bar{\psi}_\I \psi_\I \rangle_{e,\mu}$ and  $\langle \bar{\psi}_\II \psi_\II \rangle_{e,\mu}$. Note that the expectation values $\langle \bar{\psi}_\I \psi_\II \rangle_{e,\mu} = \langle \bar{\psi}_\II \psi_\I \rangle_{e,\mu}= 0$, and so they do not appear in calculations in this representation.

Another important issue to be taken into account is the relative vacuum energy associated with these. The corresponding energetics can be directly read off from the vacuum expectation values (\ref{vevHprime}) and (\ref{vevHprimeb}). Taking into account that $\sqrt{k^2 + x^2}$
is a convex function in $x$, one can use the Jensen's inequality
\begin{eqnarray}
\sqrt{k^2 + (s x + (1-s)y)^2} \ \leq \ s \sqrt{ k^2  + x^2} \ + \ (1-s) \sqrt{ k^2 + y^2}\, ,
\end{eqnarray}
valid for any $s$  between $0$ and $1$.  With this the following inequality holds:
\begin{eqnarray}
\sqrt{k^2 + m_e^2}  +   \sqrt{k^2 + m_\mu^2} \! &=& \!
\sqrt{k^2 + \left(m_1 \cos^2\bar{\theta}  + m_2 \sin^2\bar\theta \right)^2  }  +  \sqrt{k^2 + \left(m_1 \sin^2\bar{\theta}  + m_2 \cos^2\bar\theta \right)^2  }\nonumber \\[2mm]
&\leq& \! \sqrt{k^2 + m_1^2} \ + \ \sqrt{k^2 + m_2^2}\, ,
\label{ineq_b}
\end{eqnarray}
with equality valid only when $\bar{\theta} = 0$ or $\pi/2$. In deriving (\ref{ineq_b}) we have used the mass relations (\ref{mass-rel}). By inserting the previous inequality into (\ref{vevHprime}) and (\ref{vevHprimeb}) we obtain that
\begin{eqnarray}
\lan {\bar H}_0 \ran_{e,\mu} \ \geq \ \lan {\bar H}_0 \ran_{1,2}\, .
\end{eqnarray}

The fact that the representation built on $|0\ran_{1,2}$ has lower energy seems to contradict the above choice of $|0 \ran_{e,\mu}$ as the physical vacuum. In principle, this might be due  to the approximation which we have used in the Yang--Feldman equation: one could thus think that by introducing the momentum dependence via higher order corrections one would at low-enough energies observe  vacuum-energies level crossing.

Another interesting possibility would be to couple our QFT system with the general relativity (with, e.g., the Robertson--Walker geometry)
and to study in each physical phase the interplay between the QFT vacuum energy and the
gravitational energy assigned to the curvature change caused by the corresponding QFT condensate.
Such an analysis has been performed within the context of the brane-defect-filled Lorentz-Violating vacuum~\cite{sarkar},
where the extra energy of the vacuum as compared with the defect-free Minkowski vacuum has been interpreted as vacuum energy, of the type observed in the Universe today~\cite{capolupo}. Moreover, in this context, due to the Lorentz-violating properties of the microscopic ground state, the physical choice in favour of the flavor vacuum is necessitated on (broken) symmetry grounds, consistent with the findings of \cite{BlaMagueijo}. This in fact might be the key to selecting the flavour vacuum over the normal one, and may be realised, for instance, in every finite temperature situation, where Lorentz invariance is broken by the effects of the thermal bath. For early Universe (high-temperature) physics this is probably an accurate description of reality. Even in particle physics context, though, such as neutrino oscillations in the Laboratory, the absolutely zero temperature vacuum is never attained, so it is natural to accept a tiny amount of temperature present which necessitates the use of flavour vacuum for the discussion of mixing.

\section{Conclusions and perspectives}
\label{Conclusions}
\vspace{2mm}

In this paper we have analyzed  mass and mixing generation for two
fermion fields in the context of a dynamical symmetry breaking scenario.
By resorting to a formalism in which inequivalent representations of the
canonical (anti)-commutation relations are exploited for realizing the
dynamical generation of mass~\cite{UTK}, we have considered the case of
more than one generation, where the mixing terms  naturally arise.

By working in the leading order of the
Yang--Feldman expansion (corresponding to a mean-field approximation), we found that the representations in which   the  Hamiltonian
is either fully diagonal or it  contains a mixing term,  are built on unitarily inequivalent Fock spaces, thus describing different physical phases of the system.
This is quantitatively reflected in
two different sets of gap equations, which however will be discussed in detail elsewhere.

The results obtained in the simple framework here explored seem to confirm the physical relevance of the flavor vacuum, first introduced in Ref.\cite{BV95}. It is an interesting question, and object of  future work, to investigate if such features persist in the context of Higgs mechanism, and how they appear when the analysis is done using a different formalism (e.g. path-integral formulation).

\ack
\vspace{2mm}
We thank G. Vitiello and H. Kleinert for inspiring discussions related to the subject of this paper.
P.J. was supported by GA\v{C}R Grant No. P402/12/J077. The work of N.E.M. is supported in part by the London Centre for Terauniverse Studies (LCTS), using funding from the European Research Council via the Advanced Investigator Grant 267352 and by STFC (UK) under the research grant ST/J002798/1.

\section*{References}

\end{document}